\documentclass[runningheads]{svmult}

\usepackage{makeidx}   
\usepackage{graphicx}  
\usepackage{subeqnar}  
\usepackage{multicol}  
\usepackage{physprbb}  
\makeindex             

\begin{document}
\title*{Dwarf galaxies in the Local Group: the VLT perspective}
%
%
%
%
\titlerunning{The evolution of dwarf galaxies: the VLT perspective}
%
%
\author{Enrico V.Held}
\authorrunning{Enrico V. Held}
%
%
\institute{
Osservatorio Astronomico di Padova, vicolo dell'Osservatorio 5, 
35122 Padova, Italy}

\maketitle              

\begin{abstract}
%
Recent results on the evolution of Local Group dwarf galaxies obtained
from VLT imaging and spectroscopy are briefly reviewed, and prospects
for dwarf galaxy research at the VLT are discussed in the light of the
current and forthcoming instrumentation. Some aspects of future
instrument developments, such as deep wide-field imaging at both
optical and near-infrared wavelengths, that may be of advantage for
research on the evolution of dwarf galaxies, are briefly discussed.
\end{abstract}

\section{Introduction}

Understanding the origin and evolution of dwarf galaxies and their
luminosity and mass distributions may have important consequences in
modern observational cosmology. While dwarf galaxies are difficult to
study even at moderate redshifts, especially those without active star
formation, both star-forming and quiescent dwarfs can be studied with
considerable detail in the Local Group (LG).  Dwarf galaxies in
the LG are close enough that the process of star formation,
their dynamical evolution and the interplay between stars and
interstellar medium can be studied in detail.
By analyzing their color-magnitude diagrams one can derive star
formation histories and reconstruct their evolution at lookback times
comparable to the age of the universe. However, sound knowledge of the
age-metallicity relation is required to obtain reliable determination
of the star formation history.  Spectroscopy is therefore needed to
add essential information to constrain the chemical enrichment
histories.

This contribution will focus on studies of stellar populations and
kinematics of resolved Local Group galaxies.  I briefly review the
work done so far at the VLT, and discuss the prospect for dwarf galaxy
research at the VLT in the light of the existing and forthcoming
instrumentation. Some ideas for future VLT instruments are also
presented.

\section{Dwarf galaxy evolution with the VLT}
\subsection{Optical imaging}

\begin{figure}[t]
\begin{center}
\includegraphics[width=.8\textwidth]{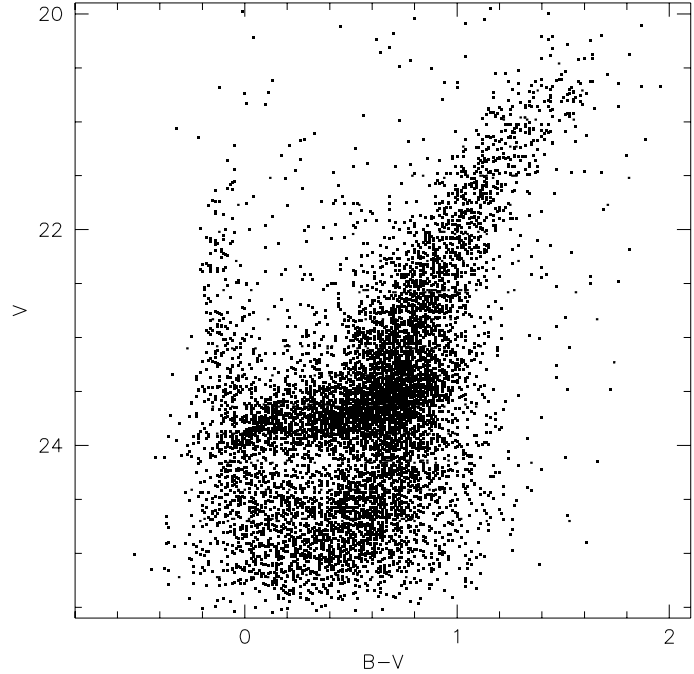}
\end{center}
\caption[]{The $V$, ($B-V$) color-magnitude diagram of the Phoenix dwarf, 
observed with FORS1 at the VLT}
\label{f_phoenix}
\end{figure}

Studies of resolved dwarf galaxies in the last decade, using both 
HST and ground based instrumentation, have provided a complex picture
of their star formation histories.  Star formation can proceed either
in distinct episodes, as in the case of Carina \cite{hurl+98}, 
or at a nearly continuous rate, as in the case of
Fornax (e.g., \cite{savi+00}, and references therein).

Deep color-magnitude diagrams of LG dwarfs obtained with FORS1/2
indicate that VLT images taken in excellent seeing can indeed be
complementary to space observations \cite{pier+99,tols+00,held+01}.
As an example, Fig.~\ref{f_phoenix} shows our observations of the
dSph/dIrr galaxy Phoenix. The main advantage of VLT imaging over
HST/WFPC2 is represented by the larger field, allowing us to detect
extended halos of red giant stars around dwarf galaxies, to trace the
populations gradients of stars in several age bins, and to map the
physical association between star formation and the interstellar
medium.
In the outer regions, the surface density of stars (``crowding'') is
relatively low, so the limiting magnitude is less subject to confusion
than in the inner regions, and the benefits of a larger collecting
power may largely balance the negative effects of a broader point
spread function.
A further advantage of a large field is the possibility to conduct
efficient searches for variable objects by using a suitable time
series strategy. This approach has being employed by our group, using
VLT and other ESO telescopes, to study the RR Lyrae variables
in LG dwarfs as tracers of the oldest stellar populations
\cite{held+01rr}.

The need for deep, wide-field optical imaging of nearby galaxies will
be largely met by VIMOS, planned for operation in 2002.  While the
advent of ACS aboard HST will re-confirm the advantage of space
observations for studying distant LG galaxies, the large field of
VIMOS will be essential to obtain deep, spatially resolved views of
the stellar populations in dwarf galaxies out to 200--300 Kpc from the
Milky Way, although it will face the competition with larger
prime-focus CCD mosaics being built at 10m-class telescopes.

\subsection{Near-infrared imaging}

\begin{figure}[t]
\begin{center}
\includegraphics[width=.8\textwidth]{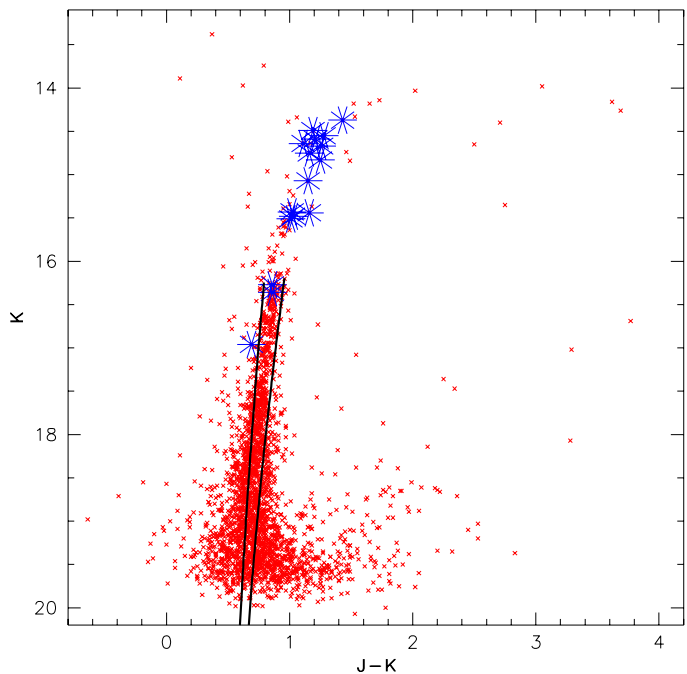}
\end{center}
\caption{The $K$, ($J-K$) near infrared 
color-magnitude diagram of the dwarf spheroidal galaxy Leo~I, 
from NTT/SOFI observations (from \cite{moma00}).
Known carbon stars from \cite{azzo+86} are also 
plotted ({\it asterisks}). Superimposed are the fiducial RGB sequences 
of the globular clusters M68 ([Fe/H]$=-2.09$) and M4 ([Fe/H]$=-1.33$)  
\cite{ferr+00} ({\it solid lines})
}
\label{f_irleo}
\end{figure}

The recent results obtained for the Magellanic Clouds and the
Sagittarius dSph using DENIS and 2MASS data have confirmed the
scientific potential of near-infrared photometry for studying the
evolved stellar populations in nearby galaxies
\cite{cion+00,niko+wein00,cole01}.
The near-infrared magnitudes and colors are more directly amenable to
the fundamental quantities -- luminosity and $T_{\rm eff}$ -- of the
stars that build up dwarf galaxies (e.g., \cite{frog+90}).  Thus,
near-infrared imaging can play an important, yet little exploited role
in studying old and intermediate age stars in LG dwarfs.
The red giant branch of Milky Way satellite dwarfs is within reach of
modern near-IR imaging detectors at 4m-class telescopes, including the
future mosaic of VISTA.  Figure~\ref{f_irleo} shows a new 
view of the RGB/AGB population in Leo~I from our NTT/SOFI survey of
evolved stellar populations in nearby dwarf spheroidals. The
use of the $K$ band allowed us to detect some very reddened luminous
stars, hidden at optical wavelengths, possibly obscured AGB stars
(cf. \cite{niko+wein00}).

On the other hand, a 10m-class telescope is needed to sample even the
relatively bright upper-AGB population in more distant LG
galaxies, and to observe red clump and subgiant branch stars in nearby
dSph. Beyond the Milky Way environment, most Local Group dwarfs
have distance moduli about 24--25 mag, which implies $K=18$--19 for
their RGB tips. Although ISAAC offers adequate sensitivity, its
small field of view is not ideally matched to the large projected
tidal radii of LG dwarfs (see \cite{mate98}).

In the next future, NIRMOS will provide imaging over a $14 \times 16$
arcmin$^2$ field in the $J$ and $H$ bands. However, the $K$
band is important in discriminating between carbon- and
oxygen-rich stars, and locate them in the theoretical HR diagram
\cite{frog+90}.
Thus, only a fully cryogenic wide-field near-IR mosaic at the VLT, with
sensitivity extended to the thermal near-IR wavelengths, would enable
deep, wide field infrared surveys of stellar populations in Local
Group galaxies, in particular of their evolved, intermediate-age
RGB/AGB populations and young red supergiants.  In that it would be
complementary to NGST, which will give superior results for the inner
regions of distant dwarfs at the edge of the LG and beyond.

\subsection{Spectroscopy}

\subsubsection{Stellar abundances}

\begin{figure}[t]
\begin{center}
\includegraphics[width=.7\textwidth,angle=-90]{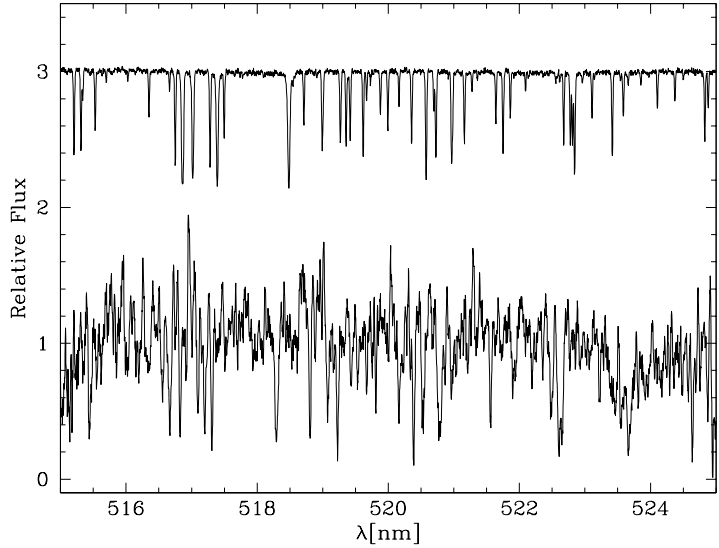}
\end{center}
\caption[]{The 50 min UVES spectrum of a red giant star in the 
Phoenix dwarf ({\it bottom}).
Also shown is the spectrum of a radial velocity standard used 
to measure the galaxy velocity by a cross-correlation technique
({\it top})
}
\label{f_uves}
\end{figure}

Tracing the chemical enrichment history of dwarf galaxies from
color-magnitude diagrams alone is a very difficult task, because of
the ambiguity between the effects of a young age and low
metallicity (the so-called ``age-metallicity degeneracy'').  Direct
abundance measurements are the best way to reconstruct the stellar
metallicity distributions in nearby galaxies and model their chemical
evolution. 

Measurements of individual stellar abundances are now
feasible at 10-m class telescopes (e.g., \cite{shet+01}).
High-resolution abundance analysis yields information on the pattern
of individual elements, which is directly related to the star
formation history of the galaxy.
%
Abundance measurements have been obtained at the VLT for a few red
giant stars in Sagittarius \cite{boni+00}, using UVES commissioning
data. Similarly to other dSph 
\cite{shet+01}, the Sagittarius dSph does not appear to be enhanced in
the $\alpha$-elements.
Luminous blue supergiants have also been observed in star-forming
dwarf galaxies out to relatively large distances. Element abundances
of O and other $\alpha$-elements, Fe-peak and s-process elements have
been measured for A-type supergiants in NGC\,6822 \cite{venn+01} using
UVES.

In the near future, FLAMES will allow us to investigate the
abundance patterns of elements in hundreds of stars in nearby dwarfs, 
although detailed abundance analysis will be possible
only for the brightest red giants in the nearby Milky Way satellites.

For this reason, 
intermediate resolution spectroscopy at the VLT will also 
play an important role in deriving metallicity distributions of
stars in  distant LG dwarfs. FORS1 spectroscopy in the Ca~II triplet
region has recently been employed to measure metallicity distributions
of stars in Sculptor, Fornax, and in the dIrr
NGC\,6822 \cite{tols+01}.  These results confirmed that the colors of
the stars are not always representative of their metal content, since
they also reflect variable Ca/Fe ratios and an age spread.

\subsubsection{Radial velocities and internal kinematics}

Precise measurements of the systemic radial velocities allow us to
investigate the dynamics and mass of the Local Group, and to establish
for some galaxies the physical association of gas and stars.  Radial
velocities of stars in the Antlia and Phoenix dwarfs have been
obtained by \cite{tols+irwi00} and \cite{gall+01}, respectively.
While low-resolution spectroscopy can give useful information, the
most compelling information on the star and gas dynamics is provided
by high resolution spectroscopy.  Recent UVES measurements of giant
stars in the Phoenix galaxy (Fig.~\ref{f_uves}) reveal that the stars
and the neutral gas have the same velocity within 2--3 {km s$^{\rm
-1}$}.


However, the most fundamental questions concern the internal 
dynamics of dSph galaxies.  What is the distribution of
mass in dwarf spheroidal galaxies ?  Is a dark halo needed to explain
the observed velocity dispersions ?  These questions not only bear on
the formation of low-mass galaxies, but also on their evolution (e.g.,
their ability to retain gas against the energetic outflows of
supernova explosions).  Many studies have been devoted to measuring
the internal velocity dispersion of dSph galaxies (e.g.,
\cite{cote+99}; see \cite{mate98} for a review). 

High resolution spectroscopy of large stellar samples in dSph's will
be a major science objective for FLAMES.  Spectroscopy out to
several core radii will be used to derive velocity dispersion profiles,
detect possible rotation, and model the mass distribution in dwarf
spheroidals.
To this purpose, a project aimed at investigating the internal
kinematics and mass-to-light ratios of nearby and distant dwarfs using
the VLT has recently been undertaken by our group.

This contribution is based on work in collaboration with I. Saviane,
Y. Momany, L. Rizzi, S. Zaggia, G. Bertelli, and G. Clementini.


\end{document}